\documentclass[12pt,a4paper,american,floatfix,pdftex,superscriptaddress,twoside,
aip,jcp,%
citeautoscript,%
prereprint,final
]{revtex4-1}%
\usepackage{amsfonts,amsmath,amssymb}
\usepackage{graphicx}
\usepackage{hyperref}
\usepackage{siunitx}
\usepackage{python}
\usepackage{makecell}
\usepackage[todos]{CMImacros}
\usepackage{lineno}

\newcommand{\cfeldesy}{\affiliation{Center for Free-Electron Laser Science, Deutsches Elektronen-Synchrotron DESY, Notkestrasse 85, 22607 Hamburg, Germany}}
\newcommand{\uhhcui}{\affiliation{Center for Ultrafast Imaging, Universität Hamburg, Luruper Chaussee 149, 22761 Hamburg, Germany}}
\newcommand{\uhhphys}{\affiliation{Department of Physics, Universität Hamburg, Luruper Chaussee 149, 22761 Hamburg, Germany}}
\newcommand{\asu}{\affiliation{Department of Physics, Arizona State University, Tempe, AZ 85287, USA}}
\newcommand{\anu}{\affiliation{Laser Physics Centre, Research School of Physics, Australian National University, Canberra, ACT 2601, Australia}}
\newcommand{\run}{\affiliation{Radboud University, Institute for Molecules and Materials, Heyendaalseweg 135 , 6525 AJ Nijmegen, Netherlands}}

\usepackage{color,soul}
\definecolor{rickscolor}{rgb}{1,0.90,0.95}
\definecolor{salahscolor}{rgb}{0.5,0.90,0.95}

\definecolor{grammarcolor}{rgb}{0,0.90,0.95}

\definecolor{qcolor}{rgb}{1,1,0}

\begin{document}
\title{Optical funnel to guide and focus virus particles for X-ray laser imaging}
\author{Salah Awel}\cfeldesy\uhhcui
\author{Sebastian~Lavin-Varela}\anu
\author{Nils Roth}\cfeldesy\uhhphys
\author{Daniel A.\ Horke}\cfeldesy\run
\author{Andrei V.\ Rode}\anu
\author{Richard A.\ Kirian}\asu
\author{Jochen Küpper}\cfeldesy\uhhcui\uhhphys
\author{Henry N.\ Chapman}\cfeldesy\uhhcui\uhhphys
\date{\today}
\maketitle



{\bfseries\noindent%
The need for precise manipulation of nanoparticles in gaseous or near-vacuum environments is encountered in many studies that include aerosol morphology, nanodroplet physics, nanoscale optomechanics, and biomolecular physics.   Photophoretic forces, whereby momentum exchange between a particle and surrounding gas is induced with optical light, were recently shown to be a robust means of trapping and manipulating nanoparticles in air.  We previously proposed a photophoretic ``optical funnel'' concept for the delivery of bioparticles to the focus of an x-ray free-electron laser (XFEL) beam for femtosecond x-ray diffractive imaging.  Here, we describe the formation of a high-aspect-ratio optical funnel and provide a first experimental demonstration of this concept by transversely compressing and concentrating a high-speed beam of aerosolized viruses by a factor of three in a low-pressure environment.  These results pave the way toward improved sample delivery efficiency for XFEL imaging experiments as well as other forms of imaging and spectroscopy.}

XFEL facilities have the capability to enable atomic-resolution images of biomolecules at physiological temperature and with time resolution down to the femtosecond regime \cite{Neutze:Nature406:752,Chapman:Nature470:73}. Since 2009 the serial femtosecond crystallography (SFX) method has yielded nearly 500 protein-structure entries in the protein data bank, many of which are from dynamic systems and are time-sequenced with time steps down to 100~fs~\cite{chapmanXRayFreeElectronLasers2019,Spence:IUCrJ4:322,Schlichting:IUCRJ2:246}.  Single-Particle Imaging (SPI) aims to enable similar capabilities, but with isolated biomolecules rather than microcrystals. Imaging single molecules at physiological temperatures would allow the observation of functional molecular motions that may otherwise be hindered in the crystal environment. Additionally, with the high data collection rates possible at XFEL facilities\cite{Ayyer:21} combined with the rapid shock-freeze method \cite{Samanta2020}, this technique might enable the detection of rare intermediate states.

The most significant present-day challenge in SPI is the production of high-density nanoparticle beams that can be directed to an x-ray beam of 100--1000~nm diameter in a low-pressure environment~\cite{Bogan:NanoLett8:310,bieleckiElectrospraySampleInjection2019,Roth:JAS124:17,Awel:OE24:6507,Kirian:SD2:041717}.  Nearly all SPI experiments have utilized aerodynamic focusing injectors for particle delivery \cite{Murphy:JAP35:1986,Liu:AST22:293, Liu:AerosolSciTech22:314}.  Such injectors are well developed and can generate particle beams with diameters on the order of 10~$\mu$m~\cite{Roth:JAS124:17,Kirian:SD2:041717,Hantke:ec5009}, but the fraction of x-ray pulses that intercept a particle still remains at less than 0.1~\%, for a 100~nm x-ray focus.  At this rate, roughly one day of continuous data collection at a 10~kHz detector frame rate would be needed for a full atomic-resolution dataset consisting of $\approx 10^6$ diffraction patterns.  For these reasons, we consider optical forces as a means to increase the target precision and density of SPI injection systems \cite{Mcgloin:FD137:335, Shvedov:PRL105:118103, Burnham:11, Shvedov:OE17:5743, Smalley2018}.  

Previously, we proposed and investigated an ``optical funnel'' that uses a focused hollow-core optical vortex beam to guide particles into a tight focus \cite{Eckerskorn:OE21:30492,Eckerskorn:PRAppl4:064001}. Our design utilized a laser that counter-propagates against the particle beam to increase the particle density both by slowing the particles as well as by forcing them closer to the beam axis. The inclusion of a surrounding gas activates photophoretic forces, caused by light absorption and subsequent momentum exchange with gas molecules, and which may be larger than optical scattering or gradient forces by orders of magnitude \cite{Shvedov:PRL105:118103}. The effectiveness of this optical funnel scheme is complicated by a number of factors that include the optical and thermal properties of the particles, radiation damage, gas pressure as limited by background X-ray scattering, and the 3D profile of the optical beam.  While many questions remain, the basic feasibility of the optical funnel concept was supported by our preliminary simulation and experimental study based on photophoretic forces that we directly measured by counterbalancing microparticles against the gravitational force \cite{Eckerskorn:PRAppl4:064001,Zhu2016,Eckerskorn:OE21:30492}.  Our previous efforts to demonstrate an optical funnel on high-speed particle beams yielded clear evidence of photophoretic forces, but no evidence of particle-beam compression, which may have been due to the rapid divergence of the optical beam and limited laser-particle interaction length.

\begin{figure*}[!ht]
	\centering
	\includegraphics[width=0.85\linewidth]{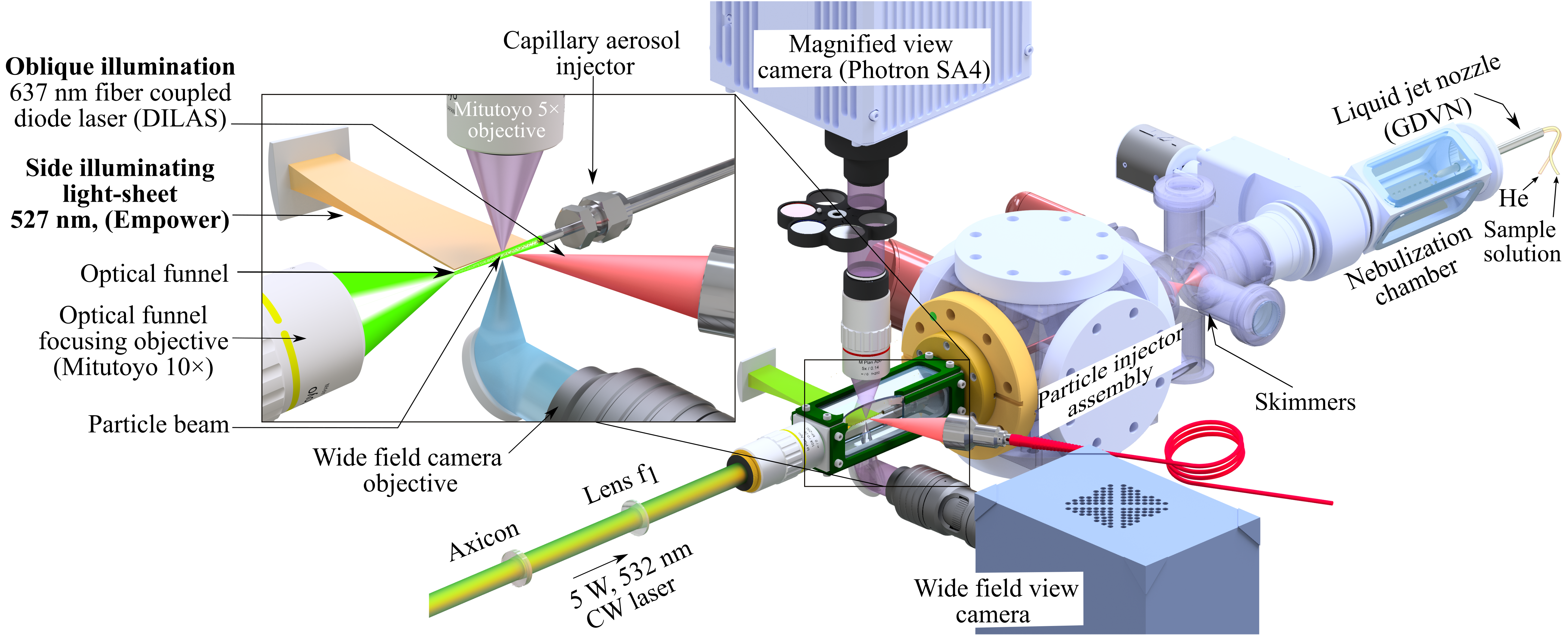}
	\caption{The basic experimental setup for aerosol particle beam imaging and focusing.}
	\label{fig:ExperimentalSetup}
\end{figure*}

\begin{figure*}[!ht]
	\centering
	\includegraphics[width=0.65\linewidth]{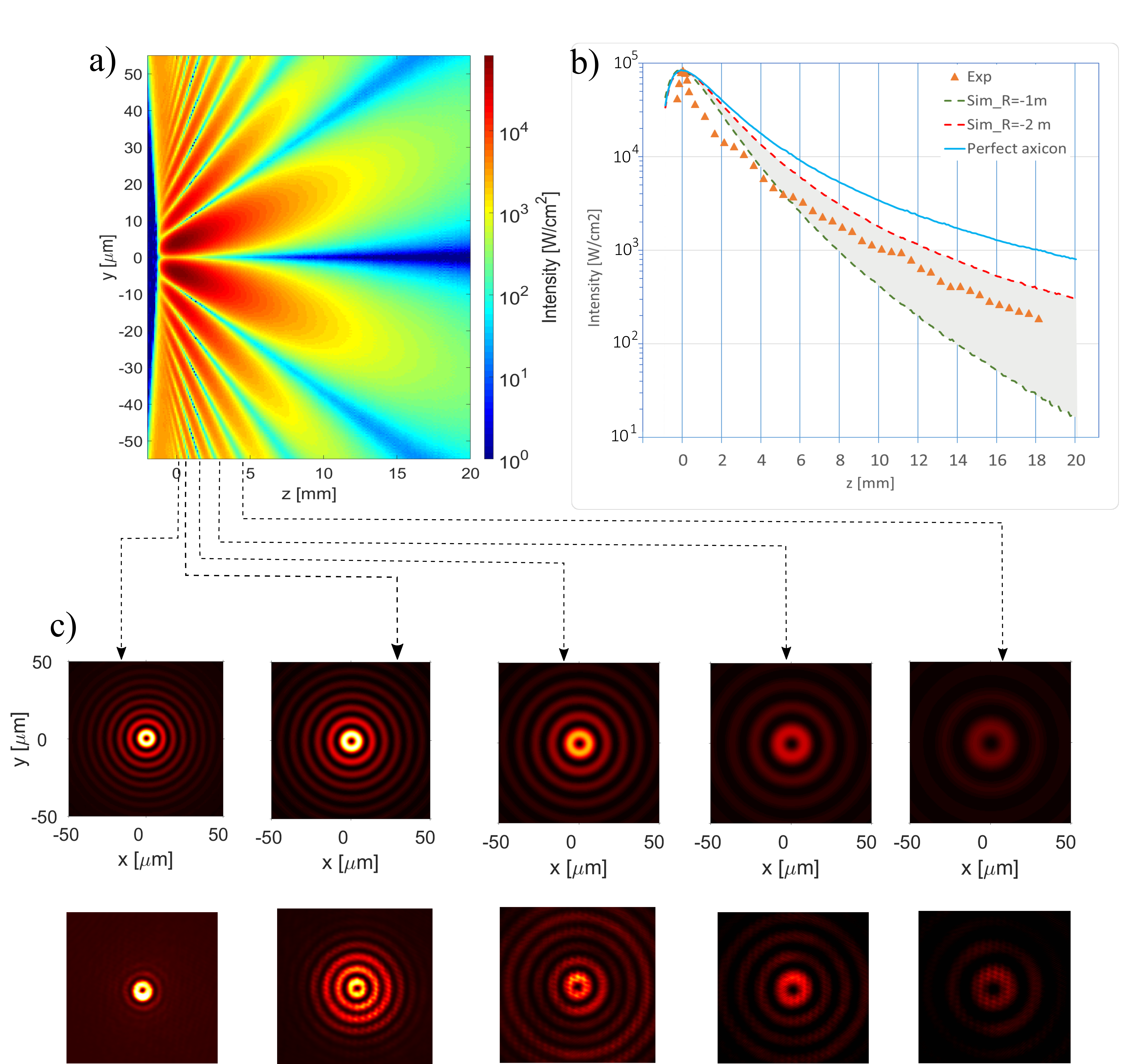}
	\caption{Slow-diverging optical funnel formed by de-magnifying a quasi-Bessel beam with a Keplerian collimator. (a) Simulated optical funnel; the distances are from the maximum intensity position. (b) Comparison between experimental intensity profile and a calculated profile in the first ring of the funnel at the total laser power of 1~W. The solid line describes a perfect axicon whereas the shaded area between the dashed lines indicate the intensity profiles for imperfect axicons with  radius of curvature within the range from $1\,\rm m$ to $2\,\rm m$ of the front-face on the axicon. (c) Beam profiles at various cross sections of the optical funnel; the top row is the results of simulations, while the bottom row are the measured profiles.}
	\label{fig:SimulatonVsMeasuredOF}
\end{figure*}

Here we demonstrate particle beam compression with an optical funnel constructed from a low-divergence hollow-core first-order Bessel beam that extends the particle--laser interaction length by a factor of up to 1000 as compared with a Gaussian beam. The basic experimental setup consists of assemblies for optical funnel beam shaping, particle injection and high-speed optical imaging is shown in figure~\ref{fig:ExperimentalSetup}. The optical funnel profile was achieved by forming a first-order quasi Bessel beam with a spiral phase plate and an axicon lens, and then re-imaging the beam inside the chamber with a de-magnifying collimator. The resulting beam changes its size due to the continuously changing magnification along the propagation direction ($\hat{z}$~axis), as shown in figure~\ref{fig:SimulatonVsMeasuredOF}. The diameter and propagation length of the optical funnel are controlled by the initial Gaussian beam diameter, the axicon geometry and refractive index, and the optical characteristics of the re-imaging collimator, namely the distance between the axicon and the collimator, the de-magnification rate, and the distance between the lenses in the collimator.  Since optimizing the optical funnel geometry for guiding a particular stream of particles is a multi-parameter task, we carried out simulations based on Fourier optics \cite{saleh2007fundamentals}, as discussed in detail in the methods section. 

Figure~\ref{fig:SimulatonVsMeasuredOF} shows a comparison of experimental laser beam profiles along with numerical simulations. For this comparison, we formed an optical funnel with a 532~nm~cw Gaussian beam with an output waist $w_0 = 1.4$~mm, which was converted to a first-order Laguerre-Gaussian vortex beam with topological charge $l = 1$ using a 16-step phase plate. The Bessel beam was formed using an axicon with a wedge angle $\alpha_0 = 0.5^\circ$. The beam was re-imaged by a de-magnifying collimator with $f_1 = 200$~mm and $f_2 = 20$~mm. The formation of this beam is illustrated in the supplementary material figure~1. The resulting optical funnel had a minimum peak-to-peak diameter of \SI{7.5}{\micro\meter} and an angle of divergence in the first bright ring of $1.3 \times 10^{-3}$~rad.  

We directed this optical funnel into a small chamber where we had prepared a counter-propagating particle beam in a low-pressure (0.4--0.9~mbar) helium gas environment as shown in figure~\ref{fig:ExperimentalSetup}. Samples of  $265 \times 265 \times 445$~nm$^3$ \textit{Cydia pomonella} granulovirus particles or 2~$\mu$m fluorescent polystyrene spheres were aerosolized with a gas-dynamic virtual nozzle \cite{DePonte:JPD41:195505} at approximately atmospheric pressure, and subsequently drawn through a gas nozzle/skimmer stage  \cite{Hantke:ec5009} in order to control the gas pressure.  A collimated beam of particles was ejected from a 2~mm inner diameter capillary in the direction opposite that of the optical beam propagation. The particles had speeds in the range of 2--20~m/s, depending on the gas differential pressures. The particle-laser interactions were observed by pulsed-laser Rayleigh-scattering imaging, which localizes the coordinates of individual particles \cite{Awel:OE24:6507,Hantke:ec5009,Kirian:SD2:041717}.  

\begin{figure}[!ht]
	\centering
	\includegraphics[width=1\linewidth]{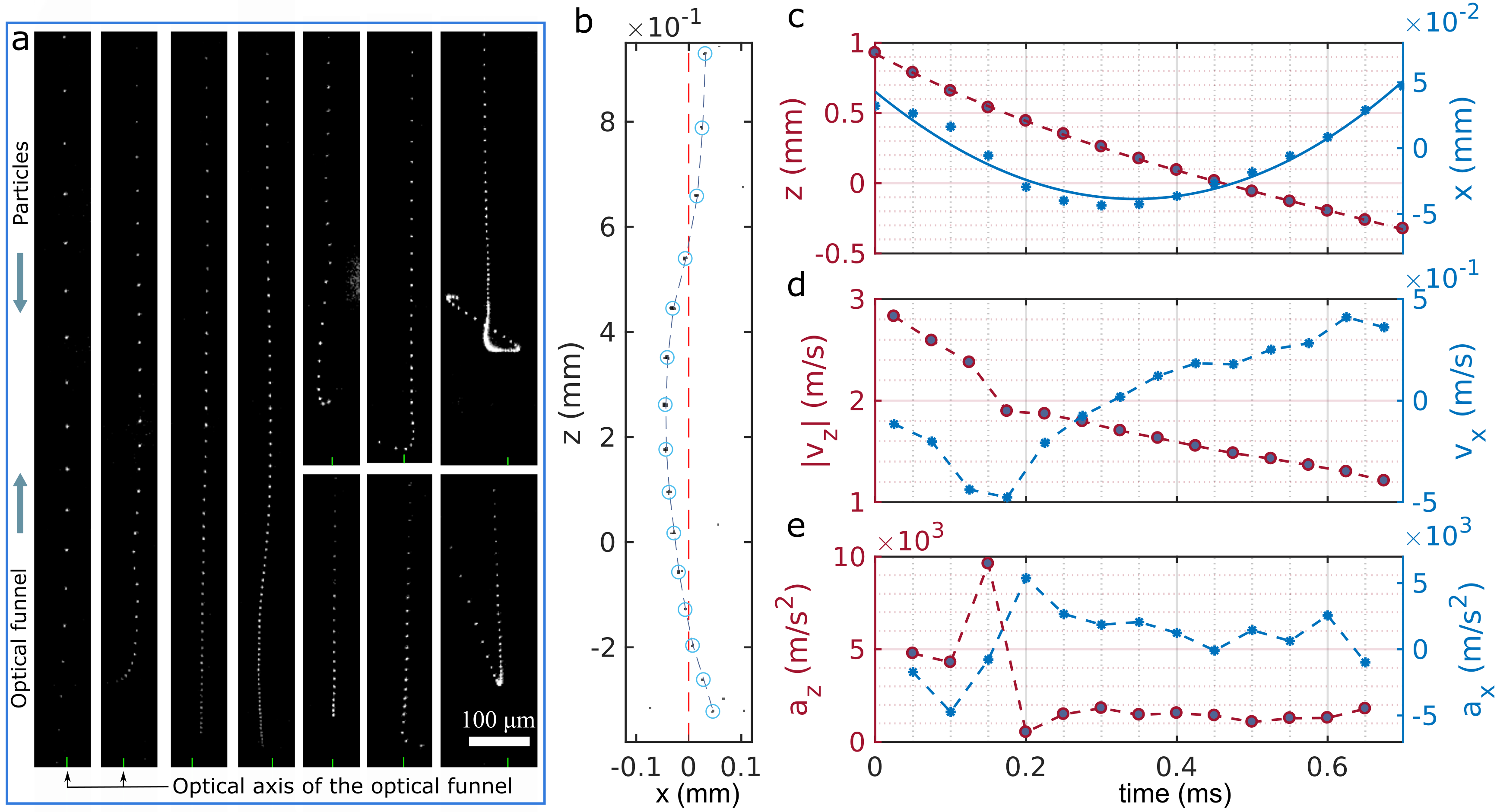}
	\caption{Granulovirus trajectories recorded with 25~kHz illumination and 0.99~mbar chamber gas pressure. (a) Background corrected raw images showing particle trajectories in a 2.0~W optical funnel. (b) Centroid positions of a single Granulovirus particle trajectory in a 0.5~W optical funnel. The optical axis of the optical funnel is indicated by the dashed red line.  (c) Calculated $x$ (blue) and $z$ (red) positions, velocities and accelerations based on particle centroids in (b).}
	\label{fig:GVDaynamics}
\end{figure}

\begin{figure*}[!ht]
	\centering
	\includegraphics[width=0.7\linewidth]{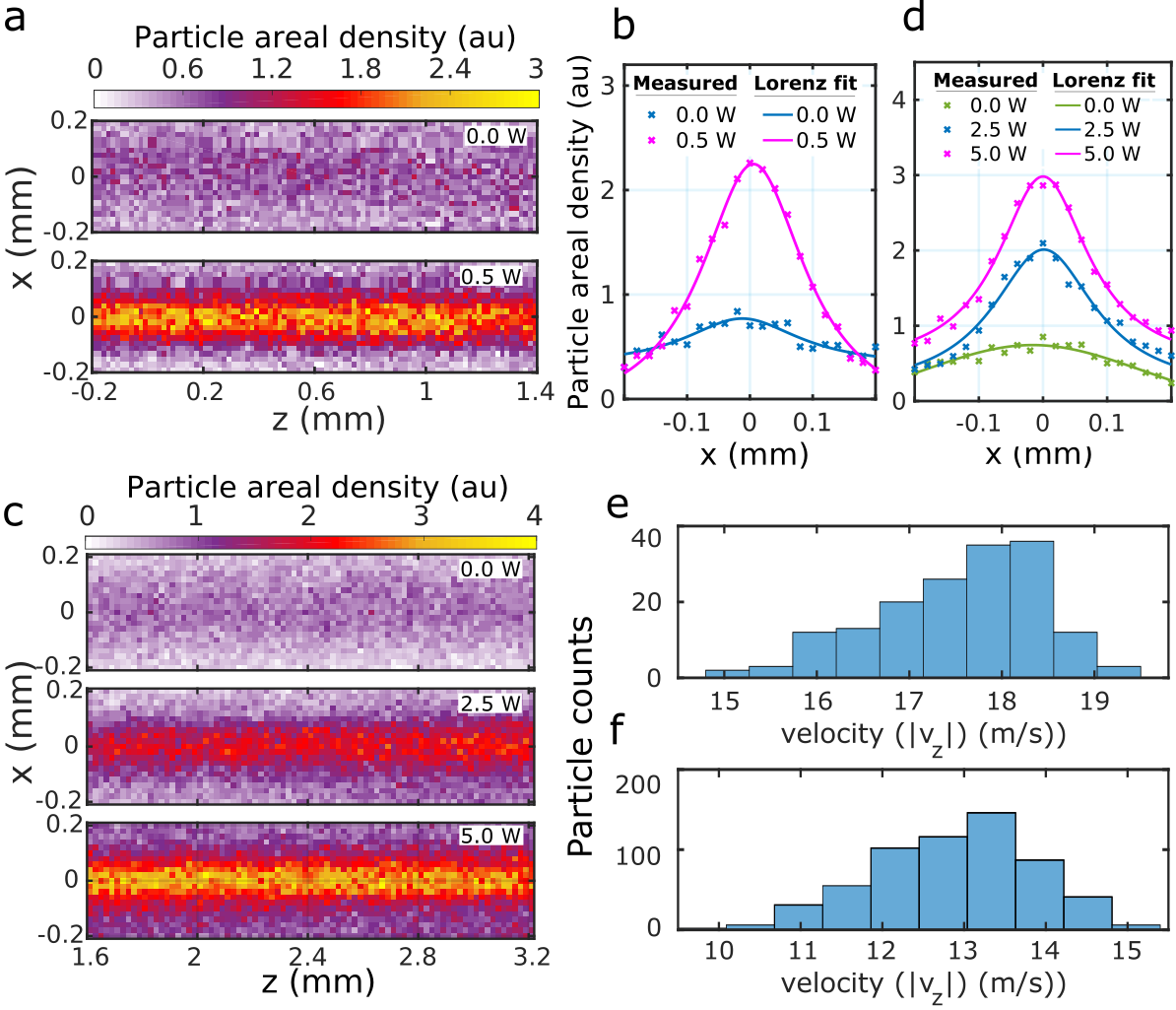}
	\caption{Focusing Granulovirus particles at 0.4~mbar chamber pressure and \SI{2}{\micro\meter}~diameter polystyrene particles at 0.5~mbar chamber pressure. (a) Granulovirus particle areal density normalized to the peak density in the laser-off. Laser-off (top) and in the presence of a 0.5~W optical funnel (bottom). (b) Lorentzian fit to the areal particle densities in (a), averaged over the $z = [0.4,0.6]$~mm region. (c) \SI{2}{\micro\meter} diameter polystyrene particles areal densities normalized to the peak density in the laser-off. Laser-off (top) and the particle beams were illuminated by 2.5~W (middle) and 5.0~W (bottom) optical funnel, respectively. (d) Lorentzian fit to the areal particle density in (c), averaged over the $z = [2.4,2.6]$~mm region. (e) and (f) are the velocity histograms of the laser-off particle densities in (a) and (c), respectively.}
	\label{fig:GVFocusing}
\end{figure*}

Figure \ref{fig:GVDaynamics} shows Rayleigh-scattering images of granulovirus particles at 0.99~mbar pressure illuminated by 637~nm, 100~ns laser flashes repeating at 25~kHz. 532~nm light from the optical funnel beam was blocked by a band-pass optical filter (Thorlabs, FL635-10) in order to isolate the Rayleigh-scattering in these images.  Multiple exposures of the same particle appear in one camera frame, which allowed us to calculate the velocities and accelerations from the particle centroid positions, as shown in figure~\ref{fig:GVDaynamics} (d,e). The granulovirus particles in figure~\ref{fig:GVDaynamics}~(b) were exposed to a 0.5~W optical funnel, with the optical beam propagating in the $+\hat{z}$ direction.  Based on the observed accelerations of 2-$10\times 10^{3}$~m/s$^2$ and estimated granulovirus mass of 2.2$\times$10$^{-14}$~g, we infer forces of 0.044--0.22~pN.  As detailed in the supplemental material, these forces correspond to a temperature difference of $\approx$~0--15.3~K across the particles. 
In absence of the optical funnel, the particle beam had a relatively well-defined velocity, for instance 17.5~$\pm$~1.1~m/s, figure \ref{fig:GVFocusing}~(e).  
However, when the optical funnel was turned on, velocities were reduced substantially; many particles stopped completely, revers their direction or deviate away from the laser, as shown in Fig.~\ref{fig:GVDaynamics}~and Fig.~s4 in the supplementary material.

Figure \ref{fig:GVFocusing}~(a) shows particle density maps that reveal the effect of a 0.5~watt optical funnel on the granulovirus particle beam at 0.4~mbar pressure.  
Panels (a) top and bottom show the 2D particle densities with the optical funnel on and off, respectively, and panel (b) shows the radial profiles of the particle beams averaged over the $z = 0.5\pm0.1$~mm region. Lorentzian fits to these profiles show that the optical funnel induces an approximately 3-fold increase in peak particle density compared with the laser-off, and the particle beam full width at half maximum (FWHM) was reduced by a factor of two. Figure \ref{fig:GVFocusing}~(c) shows density maps of \SI{2}{\micro\meter} diameter polystyrene particle beams focused by the optical funnel at 0.5~mbar chamber pressure, at laser powers of 2.5~W and 5.0~W.  Panel (c) top shows the areal particle density without the laser beam, while middle and bottom show the particle density upon optical funnel illumination at 2.5~W and 5.0~W, respectively. The focus of the optical funnel is located at the position $x=z=0$, i.e, outside the FOV of the images. The radial profiles of the particle beams, averaged over $z = [2.4, 2.6$]~mm are plotted in panel (d). The Lorentz fits show that the peak particle densities improved roughly by a factor of 3 and 5 for 2.5~W and 5~W, respectively, compared with the laser-off case. Relevant measurement conditions are listed in the supplementary material table~s1.

\begin{figure}[!ht]
	\centering
	\includegraphics[width=1\linewidth]{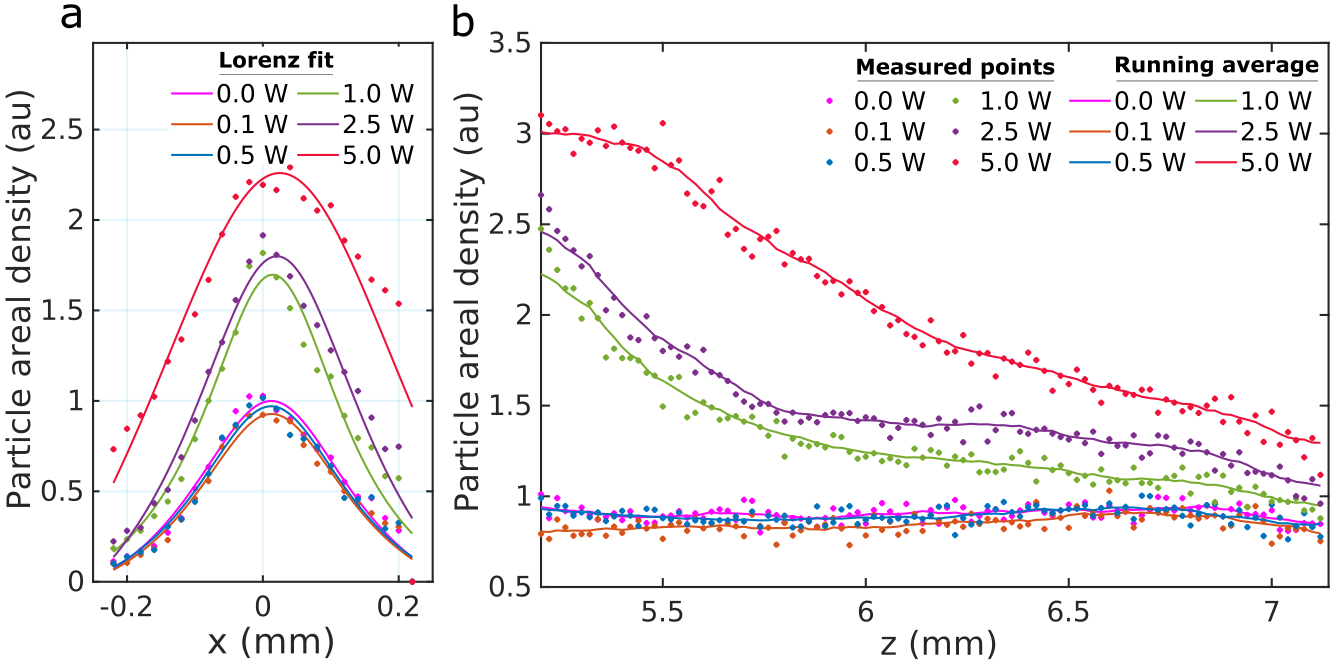}
	\caption{Increase of \SI{2}{\micro\meter} diameter polystyrene particle number density at distances between 5.2~mm and 7.2~mm before the optical funnel focus, for laser power between 0~W and 5~W.
	(a) The radial cross-section of the particle beam density at $z = 5.5$~mm from the optical funnel focus. (b) The axial cross-section along the axis of the particle beams. These number densities are normalized against the peak density for the laser-off. The solid line represents the running average calculated over 12 successive points.}
	\label{fig:2umPSFocusingDifferenLaserPower}
\end{figure}

In our observations, laser powers of the order of 1~W and beyond resulted in clearly observable changes to the particle trajectories.  
Notably, the particle beam compression effect shown in figure \ref{fig:GVFocusing} did not confine the particles to the dark 7.5~$\mu$m core of the optical funnel; the particle beam density increased within a broader $\approx 150$~$\mu$m diameter region, partly as a result of the broad (2~mm) particle beam incident on the laser.  
We also observed increases in the particle beam density at progressively larger $z$ values, i.e. the region the particles traverse before reaching the laser focus, as the laser power was increased.
Figure \ref{fig:2umPSFocusingDifferenLaserPower} shows the profiles of polystyrene particle beam densities at $z = 5.5$~mm from the optical funnel focus for five different laser powers, and for the range of $z$ values between 5.2~mm and 7.2~mm.  
The observed $z$-dependence of the particle density suggests that significant particle-laser interactions begin well before particles reach the laser focus. As particles decelerate, they interact with the laser beam for longer duration, and with sufficient laser power the particles begin to stop completely and reverse direction, as shown in supplemental material Fig.s4.  

The overall effects of the optical funnel are the result of complex, non-linear dynamics and are sensitive to several factors that include particle speed, gas pressure, laser power, and most importantly the precision of the alignment of the optical axis to the particle beam axis. While it is clear that more detailed calculations and measurements are needed in order to gain a complete understanding of the particle dynamics, our observation of overall increases in particle beam density is a significant milestone toward the development of an effective optical funnel for x-ray imaging of biological particles. Furthermore, our high-throughput manipulation of high-speed particles may stimulate developments and applications such as in attosecond dynamics, nanoscience, aerosol research, atmospheric physics, materials processing and x-ray imaging of cryo-cooled particle. Using cold particles has many fold benefits to our approach, besides enhancing the magnitude of photophoretic force (see Eq.~s13) and reducing the possible radiation damage on the optical funnel exposed particles, cooling the particle beam enables better aerodynamic focusing, especially when smaller particles are used~\cite{Samanta2020}. 

\section{Methods}

\subsection{Formation of the optical funnel}

The optical funnel was generated using a cw~532~nm Gaussian beam from a Coherent Verdi V5 diode-pumped laser.  A 16-step spiral phase plate (Holo/Or, VL-204-Q-Y-A) of topological charge $l = 1$ produced a Laguerre Gaussian vortex beam with zero-intensity on the axis and a peak-to-peak diameter of the ring, $w_{pp} = 2.42$~mm. This beam was passed through an axicon with a wedge angle $\alpha_0 = 0.5^\circ$ (Thorlabs AX2505-A) to produce a first-order Bessel beam of inner ring diameter $w_{pp}$ = \SI{80}{\micro\meter}. The optical funnel beam was produced by re-imaging the quasi-Bessel beam inside the vacuum chamber using a 1:10 demagnifying collimator, composed of a plano-convex lens ($f_1=200$~mm) and a 10$\times$ long working distance microscope objective (Mitutoyo, $f_2= 20$~mm). The optical funnel focus was formed 38~mm beyond the 10$\times$ objective, where $w_{pp}$~=~\SI{7.5}{\micro\meter}. The funnel extended $\approx$55~mm after the focus with a divergence of the central peak of $\approx$1.28~mrad. The method we used to map the profile of the optical funnel into the field of view of the camera is presented in the supplementary material section~c and Fig.~s5. 
A laser beam stabilization system (MRC Systems) was integrated in the optical path, which actively corrected the beam pointing direction with high accuracy. This kept the pointing fluctuation of the optical funnel focus position in the chamber down to $\sigma_x = 84$~nm and $\sigma_y = 72$~nm, corresponding to a beam pointing stability below \SI{4}{\micro\radian}.

\subsection{Sample preparation}
Fluorescent polystyrene spheres of $2 \pm 0.08$~$\mu$m diameter (density $\rho_\text{PS} = 1.05 ~\text{g/cm}^3$, FluoSpheres, Carboxylate-Modified Microspheres, yellow-green fluorescent) were used in these measurements. The samples were supplied as a $2~\%$ solid fraction suspended in water plus 2~mM sodium azide, which were diluted to a concentration of roughly~$1\times10^8$ particle/ml before the injection.

Granulovirus samples, which consist of 
individual viruses engulfed in crystallized protein occlusion bodies \cite{Gati:PNAS114:9}, were prepared as described previously \cite{Oberthuer:SR7:44628}. Since the majority of the granulovirus particles consist of protein crystals, we assumed its density to be similar to that of a protein crystal, $\rho_\text{GV} = 1.4~\text{g/cm}^3$ \cite{Quillin:ActaCrystD56:791}, so the mass of a single virus is estimated as 2.2$\times10^{-14}$~g.  The sample was suspended in water to a concentration of roughly 5$\times10^8$ particles/ml. An SEM image of Granulovirus particles is shown in the supplementary material Fig.~6.

\subsection{Aerosol injection}
The particles first suspended in liquid were introduced into the gas phase by injecting them into a small nebulization chamber upstream of the injector using a gas dynamic virtual nozzle (GDVN)~\cite{DePonte:JPD41:195505, Beyerlein:RSI86:125104} (see figure~\ref{fig:ExperimentalSetup}). The GDVN produced liquid droplets with diameters of approximately 1--\SI{2}{\micro\meter} with sample flow rates in the range of 1--\SI{2}{\micro\liter\per\min}. The flow rate of the GDVN helium sheath gas was in the range 10--\SI{40}{\milli\gram\per\min}. The process of aerosol formation and transporting them into the injector is described in our previous work \cite{Awel:JACR51:133}. 
Located upstream of the injector was a nozzle/skimmer stage that was used to control the pressure in the injector. It consisted of an electropolished 300~$\mu$m ID nozzle and a 500~$\mu$m ID skimmer (Beam Dynamics, Inc.) with a 2--5~mm gap between them.  The injector capillary (Swagelok \#SS-6M0-R-2) had a 2~mm inner diameter (ID) and 12.5~mm length and was mounted at the end of a 4~mm ID tube of  300~mm length, inside the main chamber as shown in figure~\ref{fig:ExperimentalSetup}.  Excess gas was evacuated from the main chamber and skimmer stage with separate scroll pumps throttled by vacuum valves.  The main particle-beam chamber and the GDVN nebulization chamber were $32  \times 32  \times 110$~mm$^3$ and $45  \times 45  \times 110$~mm$^3$ in volume, respectively (see figure~\ref{fig:ExperimentalSetup}). 

\subsection{Particle imaging}

Particles were imaged with two distinct imaging systems as shown in figure~\ref{fig:ExperimentalSetup}.  The first was a high-speed camera (Photron SA4) combined with a long working distance objective (Mitutoyo MY5X-802 -- 5$\times$) to provide a magnified field of view (FOV) of 2~$\times$~2~mm$^2$. The second was a high quantum efficiency camera (photometrics prime 95B) combined  with a variable-zoom objective (Thorlabs MVL6X12Z--6.5$\times$) and was used to produce a lower magnification with larger FOV in the range of 4~$\times$~4~mm$^2$ -- 10~$\times$~10~mm$^2$.

The illumination for both cameras was provided by either a side illuminating Nd:YLF laser (Spectra Physics Empower ICSHG-30, 527~nm, repetition rate 1~kHz, pulse duration 150~ns,  pulse energy 20~mJ, average power 20~W) or an oblique-illuminating fiber-coupled diode laser (DILAS IS21.16-LC, 637~nm, 10--100~ns pulses, repetition rates up to 1~MHz, average power 10~W). The output of the Nd:YLF laser was focused by a cylindrical lens of focal length 75~mm to form a light-sheet of size 5~mm~$\times$~20~$\mu$m parallel to and intersecting the axis of the particle beam as shown in figure~\ref{fig:ExperimentalSetup}. The intense and short pulse duration of the Nd:YLF laser produced single snapshots of many particles without significant motion blur. However, due to the relatively slow repetition rate of this illumination, images of individual particles were recorded at most once in a single frame. Therefore, such particle images were used to count particles passing through the FOV in a given time and reconstruct the two-dimensional particle density map, as discussed in section~\ref{sec:DA} below.

The output from the fiber-coupled diode laser was collimated and focused into the chamber using a fiber output collimator (Thorlabs F810SMA-635) and a plano convex lens, $f=50$~mm. Typically, the diode laser was operated at a repetition rate between 25~kHz and 100~kHz, whereas the camera recorded frames at 1~kHz and 1~ms exposure. Therefore, using this illumination, a single particle could be imaged multiple times in a single frame, as depicted in figure \ref{fig:GVDaynamics}~(a). Since the centroids of the images of particles could be accurately determined from the frames, and the time between between each snapshot was accurately known, such images could be used to extract the particle trajectories and study the particle dynamics, as discussed in section~\ref{sec:DA} below.  Importantly, due to the broad range of particle speeds and resulting overlaps in the slower-moving particle images,  it was not possible to quantitatively track all particle trajectories at a laser pulses repetition rate of 25~kHz when the optical funnel was turned on.

\subsection{Data analysis}
\label{sec:DA}
Background noise in the particle images hampers their analysis and thus must be kept to a minimum. We found that the main source of  noise in the recorded data was light scattered from the stainless steel injector tip and the chamber walls. Most of the background arising from the optical funnel  was blocked by the optical filters shown in Fig.~\ref{fig:ExperimentalSetup}, whereas the relatively constant background produced by the illumination lasers was reduced by subtracting a time-integrated median background image from every raw frame. This was followed by a spatial band-pass filtering of the images to smooth them and eliminate the remaining high frequency background.

After the pre-processing of images, particles were identified by searching for connected pixels with intensities and size above an empirically determined threshold. Intensity centroids were then calculated for each group of connected pixels for every frame, and stored as list of coordinates. For particle number density determination, centroid positions were extracted from images collected over a period of time. These positions were accumulated into the two-dimensional particle density profile as shown in  Fig~\ref{fig:GVFocusing}~(a)~and~(c). To better represent the particle density improvement by the introduction of the optical funnel we typically normalize the laser on and the laser-off 2D particle beam profiles by the maximum particle beam density measured in the laser-off. Note that in these measurements the particles were illuminated by the Nd:YLF laser. 

For velocity and acceleration analysis, first the particle trajectories must be extracted from each frame of the analysed centroid data set. This was done by searching for clusters of particle centroids in a frame which belong to the same particle trajectory, using a density-based spatial clustering of applications with noise (DBSCAN) clustering algorithm \cite{Ester96DB}. Particle velocities and accelerations were then determined from each found trajectory, using finite difference calculations based on the known illumination laser frequency, as shown in Fig~\ref{fig:GVDaynamics}~(c-e).  This characterisation of the particle dynamics can be used to calculate the forces and light-induced temperature changes on the particles, as discussed in the supplementary material. The analysis was done using a custom Matlab script.

\newpage
\section{Supplemental content}

\subsection{Modeling of the optical beam}

Our numerical optical funnel simulation is based on a free-space propagation method using Fourier optics with phase shifts caused by the optical elements \cite{saleh2007fundamentals}.  The numerical simulations start with the description of the Gaussian beam.  The electric field distribution is given, in cylindrical coordinates, by:
\begin{equation}
	E_g(r,z=0) = \Bigg(2\frac{P_{tot}}{\pi w_0^2} \Bigg)^{1/2} \exp \Bigg(-\frac{r^2}{w_0^2} \Bigg)~,
\end{equation}
where $w_0$ and $P_{tot}$ are the waist and the total power of the initial Gaussian beam, respectively, and $r = \sqrt{x^2+y^2}$ is the radial coordinate. The complex-valued field at a particular plane $z$ can be propagated by convolution with the Fresnel propagator or multiplied by the complex representation of the phase modulation of a particular optical element.
A full description of the phase modulation induced by each optical element is given below.

\textbf{Phase shift produced by a phase plate:} A vortex phase plate is a diffractive element with helical thickness variation that induces a linear phase shifts with respect to the azimuthal angle $\phi$.  This structure controls the phase of the transmitted beam azimuthally, transforming a Gaussian beam into a Laguerre-Gaussian vortex beam.  The resulting field at a distance $z_v$ from the Gaussian beam source will be
\begin{eqnarray}
	E_{l}(r,\phi,z_v) &=& E_g(r,z_v)\left(\sqrt{\frac{2}{|l|!}}\frac{r}{w_0}\right)^{|l|}\\\nonumber
	&\times&\exp\left(-\frac{r^2}{w_0^2}+il\varphi\right)
	\label{eq:EBBatZ}
\end{eqnarray} 
where $\phi$ is the azimuthal coordinate and $l$ is the azimuthal index, respectively. 
\begin{figure*}[!ht]
	\centering
	\includegraphics[width=0.85\linewidth]{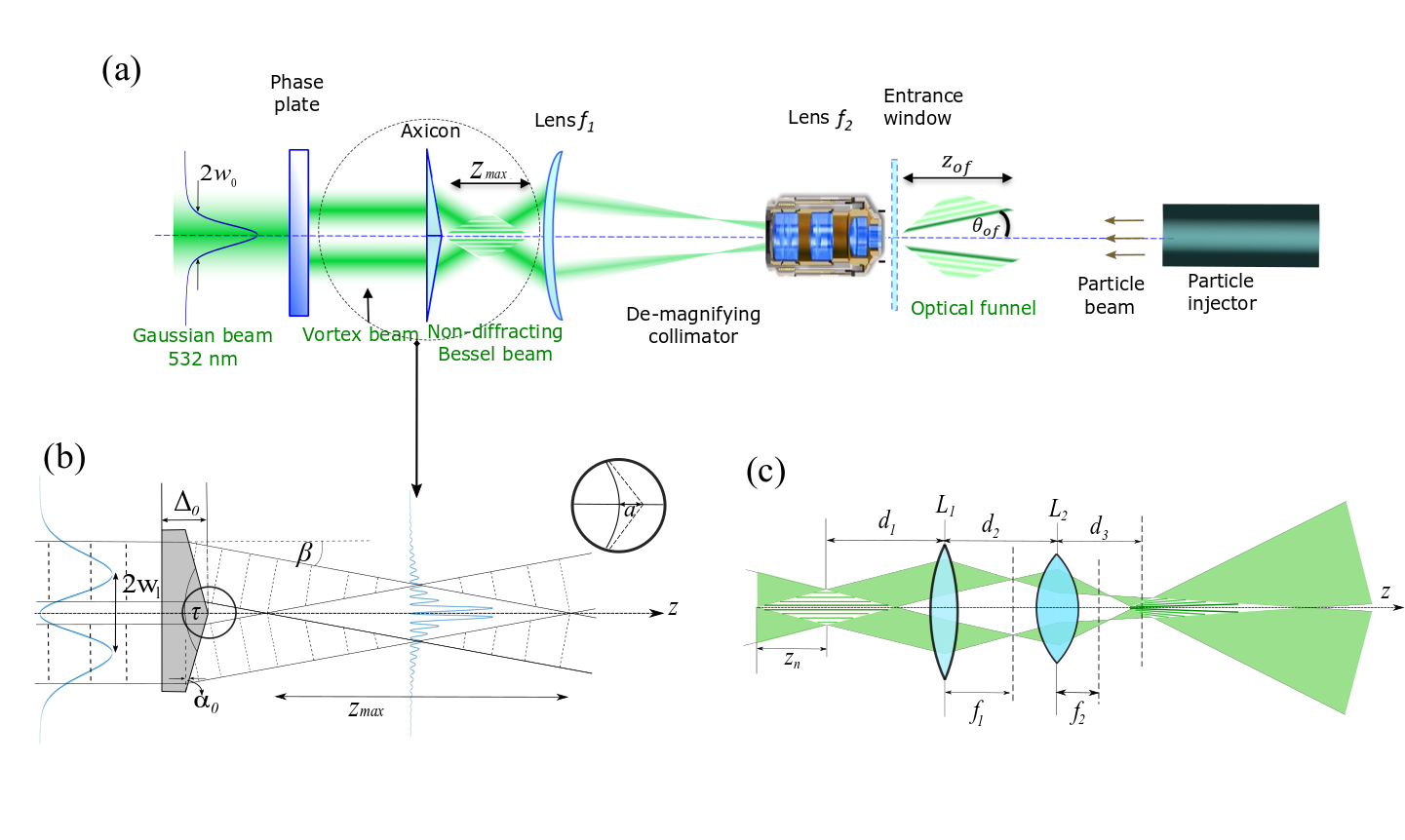}
	\caption{(a) Schematic illustration of the optical funnel formation. The initial Gaussian beam is transformed into a hollow-core Laguerre-Gaussian beam by the spiral phase plate.  The first-order Bessel beam is formed by the axicon lens.  The combined lenses $f_1$ and $f_2$ form the de-magnifying collimator that results in the slow-diverging optical funnel due to the varying magnifications of the Bessel beam image \cite{Eckerskorn:PRAppl4:064001, Eckerskorn:OE21:30492}. (b) Propagation of a high-order Laguerre-Gaussian beam through an axicon of thickness $\Delta_0$.  The plane wavefront is refracted by the axicon, forming a new beam, whose plane waves propagate over a surface of a cone with an angle $\beta$. These wavefronts interfere, creating the quasi-Bessel beam within a limited length $z_{max}$. The round-shaped tip of the axicon (inset) is included into the simulations to achieve an accurate modeling of the experimental interference pattern forming the optical funnel.(c) Constructing the optical funnel by de-magnifying a quasi-Bessel beam with a Keplerian collimator. The distances $d_1$, $d_2$, and $d_3$ are the distance from the object to first lens, the distance between lenses and the distance from the second lens to the image, respectively.}
	\label{fig:ConstructionOfOpticalFunnel}
\end{figure*}

The index $l$, also called topological charge, refers to the number of $2\pi$ cycles in the helical structure of the phase plate, making an increasing numbers of spiral staircases proportional to the index \cite{Bowman:RPP76:026401}. The topological charge is responsible for the amount of angular momentum of photons that compound the beam and the size of the hollow core of the vortex beam, both directly proportional to $l$. We note here that the diameter of the vortex beam ring at maximum intensity, $2w_l$, is related to the waist of the initial Gaussian beam by $w_l = w_0 \sqrt{| l |/2}$ .

\textbf{Phase shift produced by an axicon:} To imprint the phase shift required to transform an incoming beam into a quasi-Bessel beam, an optical element of a conical shape is used, called an axicon \cite{McLeod:54}.  When the axicon is evenly illuminated, it refracts the incoming plane waves into waves that cover a conical surface with an angle $\alpha_0$.  After passing the axicon the interfering wavefronts create an intensity profile described by several concentric rings, whose distribution depends on the topological charge of the incident beam as well as the axicon geometry and refractive index. 

The angle between the optical axis and the normal of the refracted wavefront, $\beta$, is given by axicon parameters as
\begin{equation}
	\label{eq:betaBB}
	\beta =\frac{n-n_0}{n_0}\alpha_0 = \frac{n-n_0}{n_0}\frac{\pi-\tau}{2}.
\end{equation}
for a refractive index $n$ of the axicon and the surrounding medium $n_0$, and $\tau$ is the apex angle of the axicon (Fig~\ref{fig:ConstructionOfOpticalFunnel}(b)). 

In order to compare the experimental results with the simulations we need to take into account the imperfection of the conical surface of the axicon. This was done by introducing a radius of curvature on the front-face of the axicon, which affects the intensity distribution in the propagation of the optical field. To derive an expression for the transmitted field we treat the axicon as a thin optical element. Thereby, we introduced a variable radius of curvature $R$ on the front (ideally plane) face of the axicon to find the optimal match between the simulation and experiment. The expression for the field modulation by the axicon is a modification of Brzobohaty et al.\cite{Brzobohaty:OE16:12688} who represented its conical surface as a hyperboloid of revolution of two sheets. The field is written as follows,
\begin{widetext}
	
	\begin{equation}
		\label{eq:EAxi}
		E_{ax}(r,z_{axi})=E_{l}(r,z_{axi})\exp(ikn\Delta_0)\exp\Bigg\{ ik(n_0-n)\Bigg[R\Bigg(1-\sqrt{1-\frac{r^2}{R}}\Bigg)+\Bigg(a^2+\frac{r^2}{\tan^2(\tau/2)}\Bigg)^{1/2}\Bigg]\Bigg\}
	\end{equation}
	
\end{widetext}
where $z_{axi}$ is the distance from phase plate to the axicon and $\Delta_0$ is the axicon maximum thickness. The field distribution $E_l(r,z_{axi})$ is the result of the free-space propagation over a distance $z_{axi}$ of the field produced by the phase plate $E_l(r,z_v)$ using Eq.~\ref{eq:EBBatZ}. The parameter $a$ is of least importance in our case because the vortex beam has a singularity on its axis, which leads to negligible effect due to the minimal interaction with the central region of the axicon. This can be seen when we compare the incident beam, with a waist of several millimetres, to the parameter $a$, typically in the range of tens of micrometres.  

In order to give a fully quantitative description of the Bessel beam axial propagation, it is required to provide an expression for the length and the position of focus of the beam. These expressions can be found in the work of Jarutis at al.\cite{Jarutis:OC184:105}, where they described the focusing properties of the high-order Bessel beams using an axicon, 

\begin{eqnarray}
	z_{max}&=&\frac{w_0}{\tan{\alpha_0}} \label{eq:zmax}\\ 
	z_f(l)&=&\frac{w_0\sqrt{2l+1}}{2\sin{\alpha_0}}. \label{eq:zf}
\end{eqnarray} 
$z_{max}$ in Eq.~\ref{eq:zmax} shows the maximum distance where the interference effect is still active, i.e. the limited volume where the beam is created, while Eq.~\ref{eq:zf} describes the position of maximum intensity, also called the focus of the Bessel beam. Comparing Eqs.(\ref{eq:zmax},~\ref{eq:zf}), we note that the focus position of the beam depends on the vortex order, whereas the length does not. This difference is because with the increase of topological charge, the singularity of the vortex is increased, and consequently, the interference starts at a longer distance from the axicon, whereas the region of the interference will not be modified, because the ring width does not change when the index $l$ is modified. Based on this work, we can derive an expression for the core radius of the Bessel beam when $z\gg r$:  

\begin{equation} 
	r_0=\frac{j_{(l,m)}\lambda}{2\pi\sin(\alpha_0)}
\end{equation}
where $j_{l,m}$ is the $m$-th maximum of the Bessel function of first kind and $l$-th order.

\textbf{Phase shift produced by a lens}:  After the modulation provided by the phase plate and the axicon, the beam enters the re-imaging system. A thin lens with focal length $f_i$ shapes an incoming beam by adding a phase equal to $-ikr^2/(2f_i)$. The resulting field just behind the lens is obtained from:

\begin{equation}
	\label{eq10}
	E_{Li} (r,z_{Li})=E(r,z_{Li})\exp{\left(-i\frac{k}{2f_i}r^2\right)}
\end{equation}

The distances $z_{Li}$ denote the position of the $i$-th lens relative to the previous optical element.

\textbf{Re-imaging of the Bessel beam:}  Re-imaging is important part of the optical setup as it adds the divergence to the optical funnel and controls the minimum size of the beam and its position on the axis.  The de-magnifying collimator is comprised by two thin lenses $L1$ and $L2$, with focal lengths $f_1$ and $f_2$.  The optical system can be optimised using the ABCD approach for finding the optimal distances for the desired magnification and the focus of the optical funnel.  Schematically, the re-imaging setup is shown in Fig. \ref{fig:ConstructionOfOpticalFunnel}(c).


The parameters that describe the geometry of the re-imaged beam are expressed as follows [20]:
\begin{equation}
	d_1 = \frac{(d_2f_1-f_1f_2)d_3 - d_2 f_1 f_2}{(d_2 - f_1 - f_2) d_3 - d_2 f_2 + f_1 f_2} 
	\label{eq:d1}
\end{equation}
\begin{equation}
	M(d_3) = \frac{f_2(d_2-f_1)- d_3(d_2 -f_1 -f_2)}{f_1 f_2} 
	\label{eq:M}
\end{equation}
\begin{equation}
	\Theta = \arctan{\Bigg[ \Bigg(\frac{d_2 -f_1 -f_2}{f_1 f_2}} \Bigg) r_0 \Bigg]
	\label{eq:Angle}
\end{equation}
Eq.~\eqref{eq:M} presents the magnification of the object's image depending on distance $d_3$, thus determining the radius of the optical funnel $r_0'(d_3) = M(d_3)r_0$.  It should be noted that the imaging system must satisfy the condition $d_2 > f_1+f_2$ in order to form a divergence along the propagation. The main purpose of the optimisation stage via the de-magnifying system is to re-image the Bessel beam focus into the desired spot with chosen magnification, called from now on as the focus of the optical funnel. 

The optimization procedure for the construction of optical funnels with parameters matching the experimental conditions begins taking a fixed $d_2$. Next, with the help of Eqs.~(\ref{eq:zmax}, \ref{eq:M}) we determine the desired magnification and, consequently, the focus size of the funnel.  Finally, we calculate the distance where the object to be re-imaged should be located using the Eq.~\ref{eq:d1}.  There is a last variable that will help us in the characterization of the funnels, which is the divergence of the optical funnel determined by Eq.~\ref{eq:Angle}.

\subsection{Estimated temperature gradient on the particles}
\label{sec:tempretureEstimation}

We estimate the temperature gradient across the granulovirus particles via our measured forces and the expression relating the photophoretic force to the temperature gradient in the high-Knudsen-number regime.  The granulovirus particles have a size of approximately\cite{Gati:PNAS114:9} $265 \times 265 \times 445$~nm$^3$ which is equivalent to the volume of a sphere with radius $r = 162\;\text{nm}$ and a mass of $m = 2.2\times 10^{-14}$~g.  The Knudsen number $\text{Kn} = 1320$ is defined here as the ratio of the mean-free-path of the gas to the particle radius. The mean-free-path $\lambda = 429$~$\mu$m for helium may be calculated by the formula 
\begin{align}
	\lambda = \frac{k T}{2 \sqrt{\pi} P d_m^2} 
\end{align} 
where $k$ is the Boltzmann constant, $T = 298\;\text{K}$ is the gas temperature, $P = 40\; \text{Pa}$ is the gas pressure, and $d_m = 260 \; \text{pm}$ is the kinetic diameter of helium.  The photophoretic force is derived from the gas kinetic theory\cite{Rohatschek:JAS26:717, Horvath2014} and described by the following equation: 
\begin{equation}
	F^{pp} = \frac{\pi}{6}\alpha P r^2 \frac{\Delta T}{T} 
	\label{PPFM}
\end{equation}
where $\alpha \approx 1$ is the thermal accommodation coefficient of the particle, and $\Delta T$ is the temperature difference across the particle.  The equation of motion of a particle relates acceleration to the combined photophoretic forces ($\vec{F}^{pp}$), optical scattering and absorption forces ($\vec{F}^{opt}$), and gas drag drag force ($\vec{F}^d$).  Assuming a symmetric optical beam, we expect two relevant force components for a particle in the $x$--$z$ plane:
\begin{equation}
	\Bigg\{
	\begin{aligned}
		m a_x &= F^{pp}_x + F^{opt}_x+ F^d_x \\
		m a_z &= F^{pp}_z +F^{opt}_z+ F^d_z
	\end{aligned}
	\label{eq:EquationOfMotion}
\end{equation}
where the subscripts $x$ and $z$ refer to the transverse and axial components, respectively.  The slip-corrected drag force $\vec{F}^{d}$ for a particle in the high-Kn regime is
\begin{equation}
	\begin{aligned}
		{\vec{F}^{d}} &= {\frac{6\pi\mu r (\vec{v}_g-\vec{v}_p)}{C}}\\
		C   &= 1+  \mathrm {Kn} (c_{1}+c_{2}\cdot e^{\frac {-c_3}{ \mathrm {Kn} }}), 
	\end{aligned}
	\label{eq:StocksCorrected}
\end{equation}
where $C$ is the Cunningham slip-correction factor, introduced by Knudsen and Weber, with empirical coefficient of, $c_1 = 1.231$, $c_2 = 0.4695$ and $c_3 = 1.1783$ 
\cite{Roth:JAS124:17, Hutchins:AST22:202}. $v_p$ and $v_g$ are the velocity of the particle and the gas, respectively. 

\begin{figure}[h!]
	\centering
	\includegraphics[width=1\linewidth]{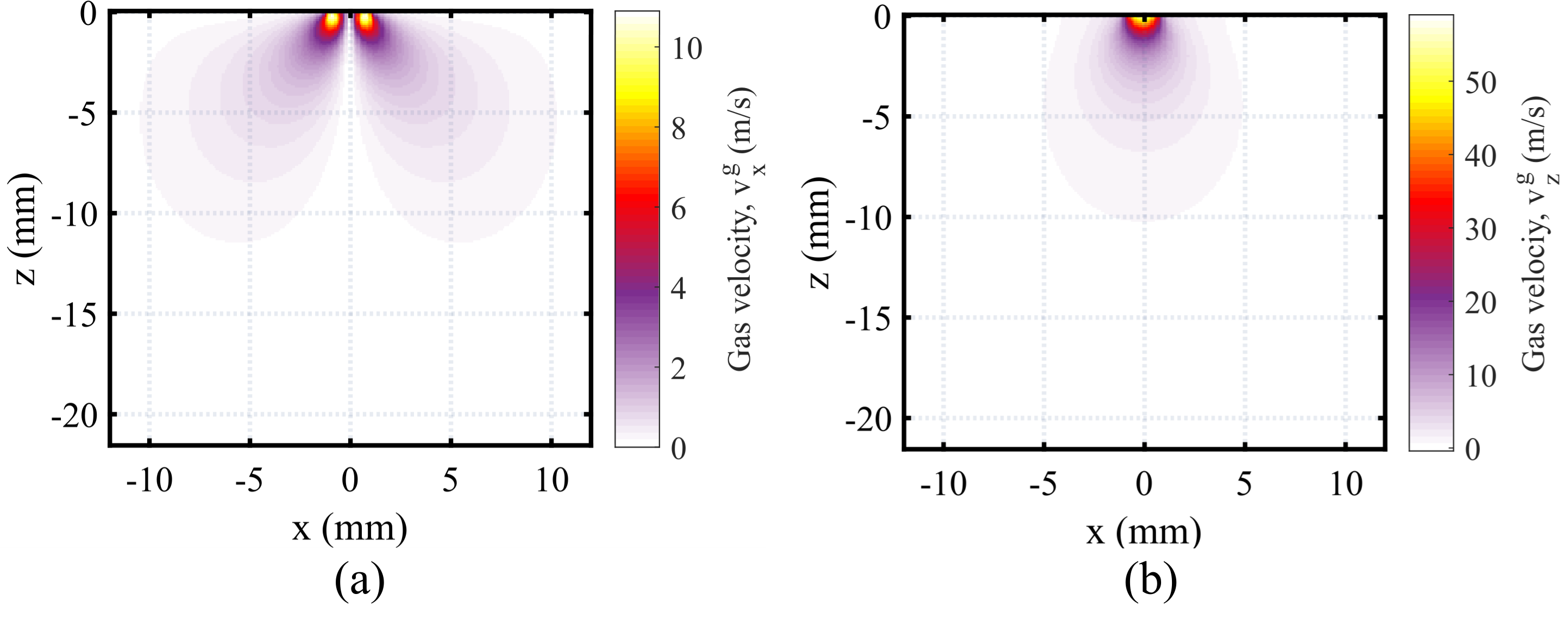}
	\caption{The simulated He gas velocity between the exit of the aerosol injector and optical funnel focus at 0.99 mbar chamber pressure.(a) The velocity along the transverse direction and (b) the velocity along the axial direction. The injector exit is at $z=0$. This simulations were performed as described in our previous publication [\onlinecite{Roth:JAS124:17}].}
	\label{fig:Vg}
\end{figure}

\begin{figure*}[!ht]
	\centering
	\includegraphics[width=0.85\linewidth]{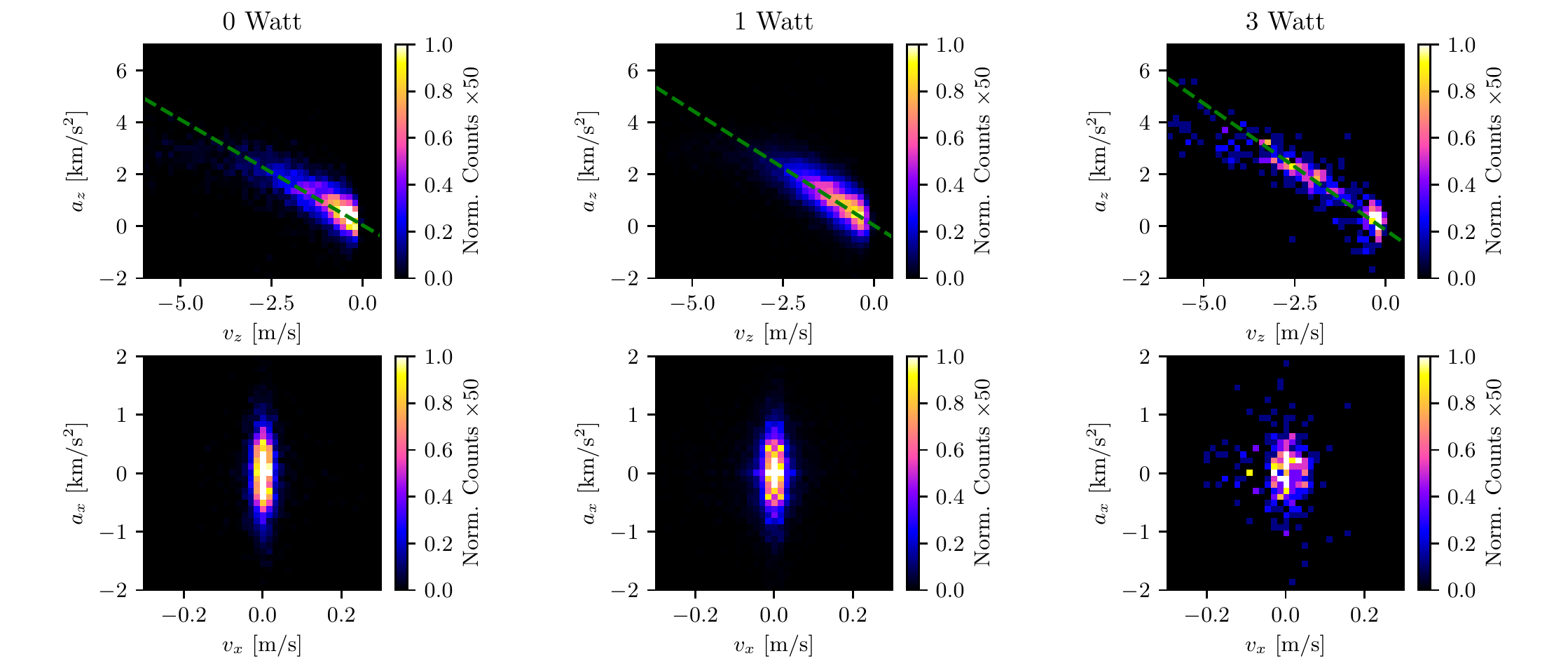}
	\caption{Granulovirus velocities and accelerations under different optical funnel illumination conditions, each at 0.99~mbar pressure (see table~\ref{table:Gvtrajectories} for experimental parameters).  Histograms of particle acceleration and velocity for the $z$ and $x$ axes are shown in the top and bottom rows, respectively, with columns corresponding to 0, 1 and 3 Watt optical funnel powers, respectively.  The velocities and accelerations were estimated from finite differences in centroid positions using triplets of sequential images determined from 25~kHz stroboscopic images.  The green dashed line in the middle row corresponds to a linear fit to the function $a_z = a v_z + b$.  The resulting values for the $\{0, 1, 3\}$-Watt optical funnel illumination were $a=\{-810, -890, -980\}$~s$^{-1}$ and $b=\{0.043, 0.030, -0.15\}$~km/s$^2$.}
	\label{fig:drag-forces}
\end{figure*}

In order to understand the importance of gas drag forces, we numerically simulated the gas velocity using the injector geometry and the measured pressures upstream and downstream of the injector as an input. The Navier--Stokes equations were solved with a finite-element solver using the COMSOL Multiphysics software suite \cite{Roth:JAS124:17}. As shown in figure \ref{fig:Vg}, close to the exit of the injector the gas expands with relatively high velocity, which also accelerates the particles. However, below 15~mm from the the injector, where our measurements were performed, the gas decelerates to negligible velocity.  Under the assumption of negligible gas velocity, equation \ref{eq:StocksCorrected} predicts a deceleration of approximately $2.2\times10^4$~m/s$^2$ for granulovirus particles in absence of laser illumination at the typical measured speed 17.5~m/s.  This predicted value is not far from our measured acceleration of $1.42\times 10^4$~m/s$^2$, which we obtained from the acceleration/velocity histograms shown in Figure~\ref{fig:drag-forces}.  These histograms moreover reveal a relatively small acceleration of $a_z = 43$~m/s$^2$ at zero velocity.  Based on these calculations and observations, we assume the gas velocity to be zero in subsequent calculations.


We now consider the particle trajectory shown in main text figure~3, in which the peak acceleration at the time $t=0.15$~ms is $9.6\times 10^3$~m/s$^2$ at a particle speed of 2.2~m/s. At 0.99~mbar, the mean-free-path of helium gas is \SI{172}{\micro\meter}, and Knudsen number, $\mathrm {Kn} = 528$, the relevant measurement parameters are listed in table~\ref{table:Gvtrajectories}. After subtracting the slip-corrected drag force, the combined photophoretic and optical force is $6.2\times 10^{-14}$~N, which suggests that the photophoretic force is dominant since the maximum possible radiation pressure is $F_z^{opt} \approx \pi r^2 I/c \approx 8.3\times 10^{-14}$~N at the peak intensity in the optical funnel ($I = 3.0\times 10^8$~W/m$^2$).  Thus the peak acceleration suggests that $\Delta T/T = 0.045$, and $\Delta T \le 13.4$~K since $T \ge 298$~K. If we assume the thermal conductivity of granulovirus to be $g = 0.3$~W/m/K (comparable to most polymers and protein-based materials) then the approximate intensity of thermal energy transferred across the granulovirus at this temperature difference is of order $I_\text{heat} \approx g \Delta T/ r \approx 2.47\times 10^7$~W/m$^2$.

\begin{figure*}[!ht]
	\centering
	\includegraphics[width=0.75\linewidth]{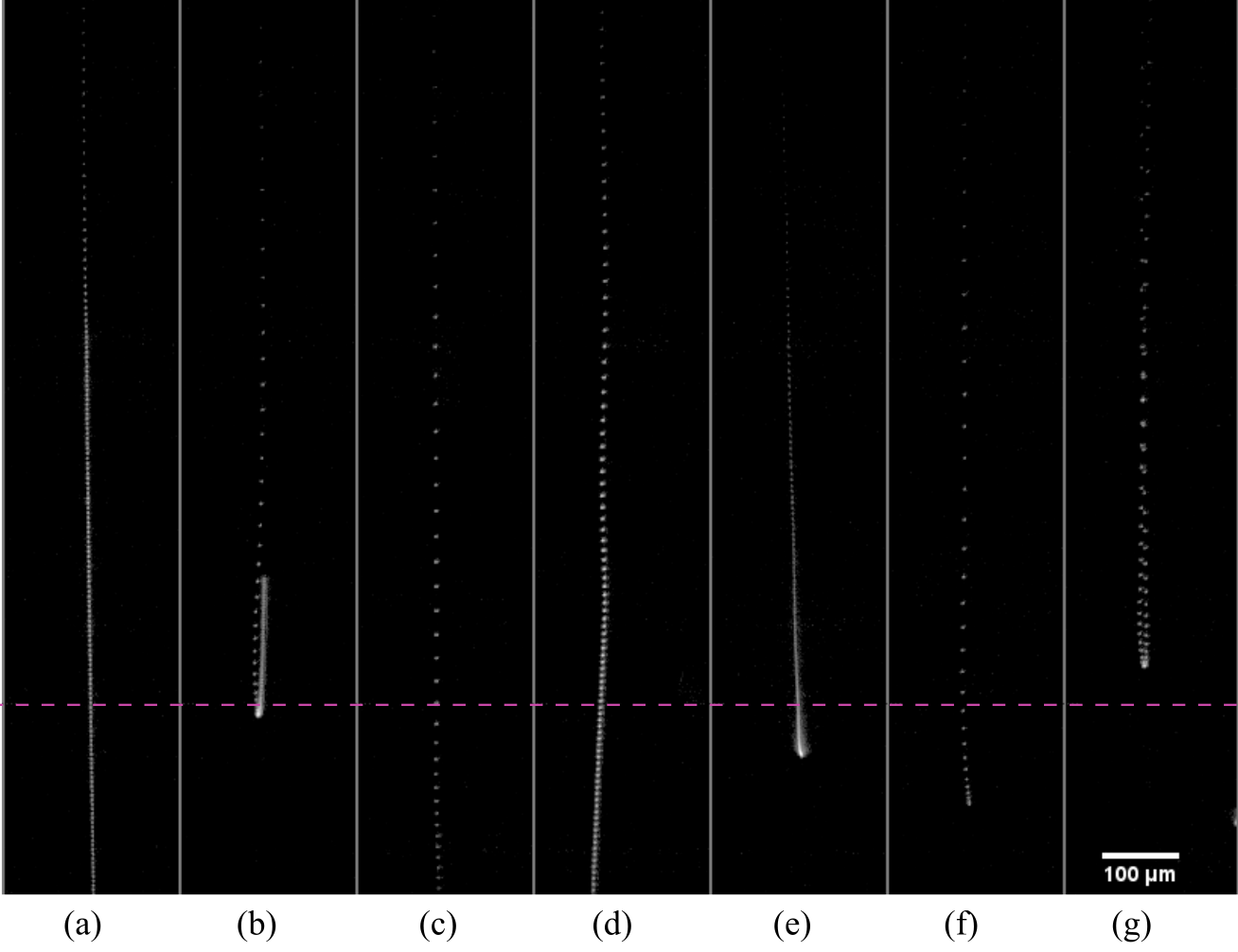}
	\caption{Selected background corrected stroboscopic images of 2~micrometer diameter polystyrene particles trajectories in 5~W optical funnel. This measurement was performed at 1.5~mbar chamber pressure and the particles were illuminated by the DILAS laser operating at 100~kHz. In this images, the optical funnel was propagating in the vertical direction and  the horizontal dashed-line indicates the position of focal plane in the images, how we determined this position is shown in Fig.~\ref{fig:OpticalFunnelProfile}.}
	\label{fig:2umParticlesTrajectoreis}
\end{figure*}

\subsection{Mapping the position of the optical funnel in the camera FOV}
\label{sec:MappingOfTheopticalFunnel}
To map the location of the optical funnel in the camera FOV we rely on Rayleigh scattering imaging of high density particles illuminated by the optical funnel. In these measurements, the optical funnel was propagating opposite to the particles and the scattered intensity was recorded on the camera which was set to a relatively long exposure time of 20~ms. This produce long streak-images of the particles, similar to streak-imaging described in \cite{Awel:OE24:6507}. These images were pre-processed to remove the background and the scattered image intensities were integrated through the measured frames. Then the beam profile is reconstructed by pixel by pixel weighted averaging of this integrated image intensity. 
\begin{figure}[!ht]
	\centering
	\includegraphics[width=0.95\linewidth]{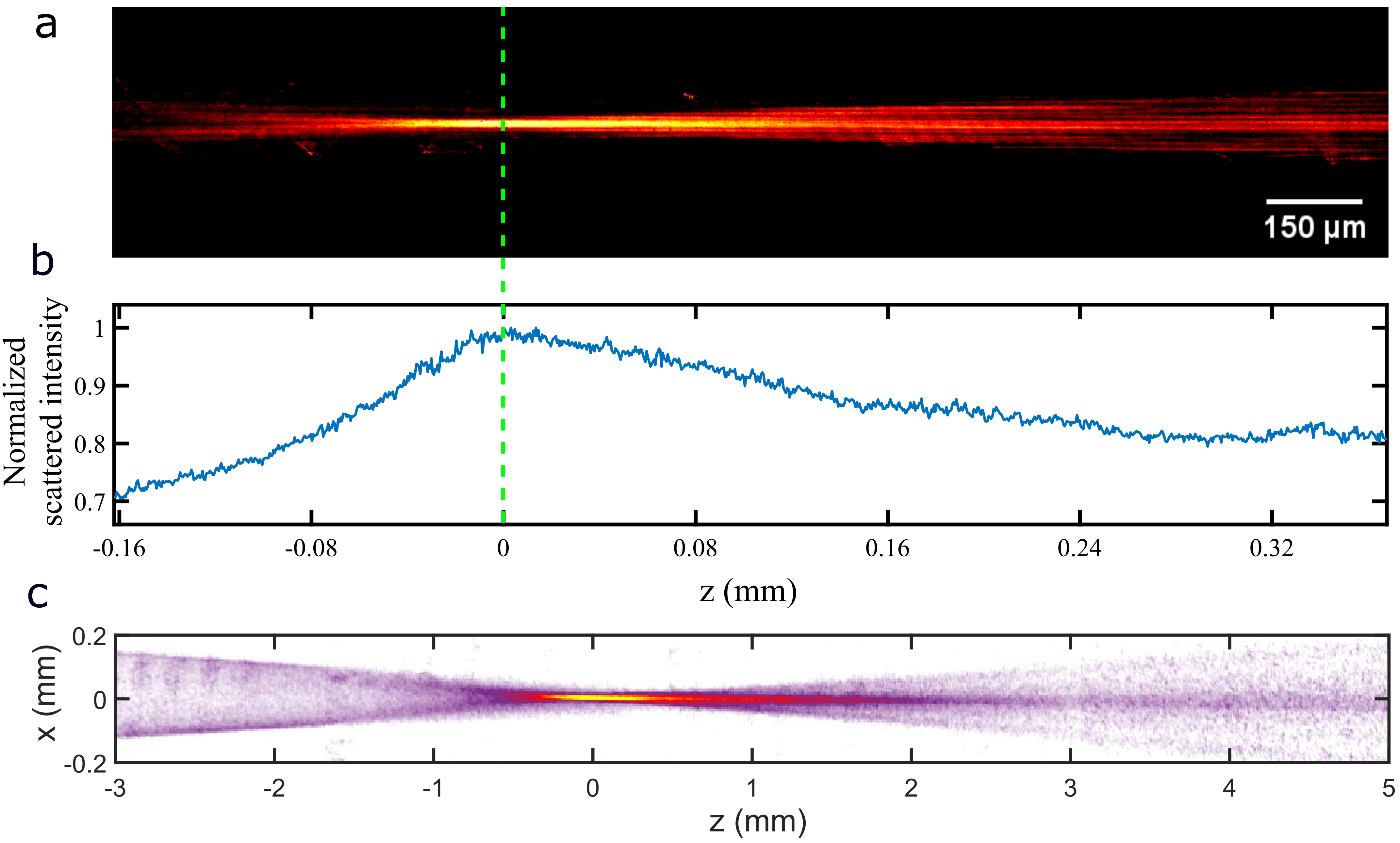}
	\caption{Mapping the optical funnel in the camera FOV using particles or air scattering intensities. (a) The reconstructed two dimensional intensity profile of the optical funnel. This profile is generated by integrating 5000 frames, each containing long exposure scatted intensity of GV particles illuminated by a 0.2 watt optical funnel. (b) Normalized axial cross-section of the intensity profile in (a). The green dashed-line indicates the focal plane of the optical funnel. (c) The optical funnel profile generated by time integrating air scattering in the beam path.}
	\label{fig:OpticalFunnelProfile}
\end{figure}

Figure~\ref{fig:OpticalFunnelProfile}~(a) shows the two dimensional reconstructed optical funnel profile generated from the 5000 frames each containing streaked scattering intensity of granulovirus particles illuminated by a 0.2~watt optical funnel. The axial cross-section of the beam is shown in figure~\ref{fig:OpticalFunnelProfile}~(b). The maximum intensity position, indicated by the green dashed line, shows the focus of the optical funnel. This position is used to define the center of our coordinate system. Similarly, the beam position can also be inferred by time integrating air scattering intensity in the optical funnel path as seen in Fig~\ref{fig:OpticalFunnelProfile}~(c). Using these techniques, we can determine the focal position of the optical funnel with a resolution better than a micrometer.

\begin{figure}[!ht]
	\centering
	\includegraphics[width = 0.55\linewidth]{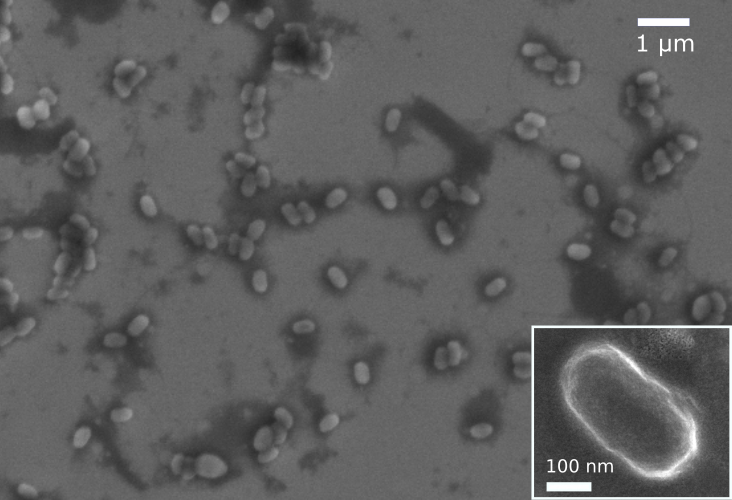}
	\caption{Scanning electron microscope image of granulovirus particles suspended on silicon substrate.}
	\label{fig:GV-SEM}
\end{figure}
\newpage

\section{Measurement parameters}
\begin{table*}[!ht]
	\centering
	\begin{tabular}{  p{6.2cm}  p{3.6cm}  p{3.6cm}}	
		&\textbf{granulovirus} 					    			 		& \textbf{2~$\mu$m polystyrene}\\  \hline
		Mean particle velocity, $V_z$, with laser-off (m/s)    &  17.4~$\pm$~0.93 										& 12.87~$\pm$~0.97\\
		Optical funnel power (W)                 			   	&0, 0.5  		              						&0, 2.5, 5.0\\ 
		Chamber pressure (mbar)          		& 0.4     							         &0.5\\ 
		Sample concentration (particles/ml)  		&5$\times10^8$                            &7.5$\times10^7$  \\
		\makecell[l]{Particle~generation~rate \\ (particles/second)}         &$1.17\times10^4$          & $2.25\times10^3$    \\
		GDVN gas flow rate   								  & 15~mg/min                             & 16~mg/min \\		
		\hline
	\end{tabular}
	\caption{Experimental parameters for the particle beam focusing presented in Fig.~4 in the main text.}
	\label{table:FocsusingParticleBeam}
\end{table*}

\begin{table*}
	\centering
	\begin{tabular}{  p{8cm}  p{3.6cm}  p{3.6cm}}	
		&\textbf{granulovirus} 					    			 		\\   \hline
		Mean particle velocity, $V_z$, with laser-off (m/s)    &  -1.72$\pm$~0.42 									\\
		Optical funnel power (W)                 			   	&0, 1.0, 3.0  		              						\\ 
		Chamber pressure (mbar)          		& 0.99     							         \\ 
		Sample concentration (particles/ml)  		&5$\times10^8$                             \\
		\makecell[l]{Particle~generation~rate \\ (particles/second)}         &$2.18\times10^4$         \\
		GDVN gas flow rate   								  & 14~mg/min                             \\		
		\hline
	\end{tabular}
	\caption{Experimental parameters for the granulovirus particle dynamics presented in Fig.~3 main text and Fig.~\ref{fig:drag-forces}.}
	\label{table:Gvtrajectories}
\end{table*}

\newpage
\section{Acknowledgments}
In addition to DESY, this work has been supported by the Clusters of Excellence at Universit{\"a}t Hamburg, the “Center for Ultrafast Imaging” (CUI, EXC 1074, ID 194651731), and “Advanced Imaging of Matter” (AIM, EXC 2056, ID 390715994) of the Deutsche Forschungsgemeinschaft (DFG), by the European Research Council through the Consolidator Grant COMOTION (ERC-614507) and by the Australian Research Council's Discovery
Projects funding scheme (DP170100131). R.A.K.\ acknowledges support from an NSF STC Award (1231306).

\section{Author contributions}
S.A., R.A.K., D.A.H., A.V.R., J.K., and H.N.C. conceived the idea and designed the experiment. S.A., D.A.H., N.R., R.A.K. and A.V.R. performed the measurements. S.A. and R.A.K. developed the particle data analysis tool and analyzed the data. S.L.V. and A.V.R. performed the optical beam modeling. S.A. wrote the initial version of the manuscript with inputs from R.A.K., S.L.V. and A.V.R. All authors contributed in  scientific discussions and  manuscript revisions. 


\setlength{\bibsep}{2pt plus 0.3ex}
\renewcommand{\bibsection}{\section*{References}}
\renewcommand*{\bibfont}{\normalfont\small}
\bibliography{funnel}
\end{document}